\documentclass[aps,prc,floatfix,groupedaddress,showpacs,amsfonts,nofootinbib,twocolumn]{revtex4}
\usepackage{bm}
\usepackage{epsfig}
\usepackage{amsmath}
\usepackage{dcolumn}
\usepackage{slashed}

\begin{document}

\title{Relativistic R matrix and continuum shell model}

\author{J. Grineviciute}
\affiliation{Department of Physics, Western Michigan University, Kalamazoo, MI 49008}

\author{Dean Halderson}
\affiliation{Department of Physics, Western Michigan University, Kalamazoo, MI 49008}

\begin{abstract}
\begin{description}
\item[Background] The $R$ matrix formalism of Lane and Thomas has proven to be a convenient reaction theory for solving many-coupled channel systems.  The theory provides solutions to bound states, scattering states, and resonances for microscopic models in one formalism.

\item[Purpose] The first purpose is to extend this formalism to the relativistic case so that the many-coupled channels problem may be solved for systems in which binary breakup channels satisfy a relative Dirac equation.  The second purpose is to employ this formalism in a relativistic continuum shell model.

\item[Methods] Expressions for the collision matrix and the scattering amplitude, from which observables may be calculated, are derived.  The formalism is applied to the 1p-1h relativistic continuum shell model with an interaction extracted from relativistic mean-field theory.

\item[Results]  The simplest of the $\sigma +\omega +\rho$ exchange interactions produces a good description of the single-particle energies in $^{16}$O and $^{90}$Zr and a reasonable description of proton scattering from $^{15}$N.

\item[Conclusions] The development of a calculable, relativistic $R$ matrix and its implementation in a $1p-1h$ relativistic continuum shell model provide a simple
 relatively self-consist, physically justifiable model for use in knockout reactions.
\end{description}
\end{abstract}


\maketitle

\section{Introduction}

The R matrix formalism of Lane and Thomas \cite{LaTh58} has proven to be the most physical and convenient reaction theory for solving many-coupled channel systems in light and medium mass nuclei.  It is not uncommon to couple thirty or more residual states of the target in non-relativistic calculations, and new computer codes need not be written each time states are added.  Microscopic models and non-local potentials are easily incorporated in the theory.  In addition to providing scattering states, the formalism yields bound states and resonances.  Coupled-channels techniques which involve integrating coupled differential equations can become unstable for large numbers of channels, and they can miss narrow resonances because the equations must be solved for each energy over the resonance.  Also, scattering observables are calculated quickly at a given energy in the R matrix formalism because they require diagonalizing matrices whose dimensions are just the number of channels.  Additional advantages may be found in a review article by Descouvemont and Baye \cite{De10} and applications in a review in Ref. \cite{Ha05}.

This paper is the last of a series of three articles that describe the extension of the $R$ matrix theory to the relativistic case so that the many-coupled channels problem may be solved for systems in which binary breakup channels satisfy a relative Dirac equation.  The first paper \cite{Ha88} demonstrated that an $R$ matrix theory exists for the Dirac equation and derived the appropriate Bloch operator.  Then an example was given for 35.5 MeV neutron scattering from a Woods-Saxon potential.  The expansion basis consisted of the free-particle Dirac solutions whose upper components were zero at twice the $R$ matrix radius.  

Paper \cite{Ha09} demonstrated that Dirac oscillator wave functions \cite{IM67,M89} provided an excellent and convenient expansion basis.  This article also demonstrated that the $R$ matrix formalism allows one to easily orthogonalize scattering solutions to bound-state solutions and to treat non-local potentials; and, hence, to calculate exchange terms in relativistic impulse approximation exactly.  Examples were given for 160-200 MeV elastic proton scattering from $^{16}$O, $^{40}$Ca, and $^{90}$Zr in the impulse approximation with the two-nucleon $t$-matrix elements of Ref. \cite{H85}.  In Ref. \cite{Ha09}, it was shown that the common local-density approximation for the exchange terms was inadequate in relativistic calculations.  The discrepancy between the exact and local density approximation calculations was traced to the extreme difference between the matrix elements of the negative energy states of the basis functions, and, hence, was a relativistic effect.

The present paper provides derivations of the collision matrix expression for coupled channels and the scattering amplitude from which scattering observables can be extracted.  As an example of the formalism, relativistic continuum Tamm-Dancoff approximation (TDA) calculations for 16O are performed with interactions derived from relativistic mean-field theory.  Specifically, the formalism referred to as quantum hydrodynamics (QHD) \cite{SW86} is employed.  The classical meson fields of the original QHD are replaced by one-meson exchange potentials.  The validity of this replacement is checked by comparing single particle energies (SPEs) for $^{90}$Zr, calculated from both treatments with the same coupling constants.  Surprising agreement is found between the two procedures with the simple  $\sigma +\omega +\rho$ exchange.  In addition the simple  $\sigma +\omega +\rho$ exchange with QHD coupling constants provides reasonable agreement with experimental $^{15}$N$(p,p)^{15}$N cross sections at 39.84 MeV.  This is, therefore, a simple, physically justifiable interaction for later use in knockout reactions.  The importance of coupled-channels solutions in $(e,e^{\prime}x)$ was emphasized in Ref. \cite{Ha96}. Finally, the role of pions is investigated.  It is found that pions have a significant effect on SPEs and the $^{15}$N + $p$ cross section, however, a definitive conclusion on their utility awaits a better approximation for the matrix elements with pseudovector $\pi N$ coupling.

\section{$R$ matrix formalism}

Solutions to the one-channel Dirac equation will be written in the two-component form
\begin{equation} 
\label{Eq1}
u_{D} =
\begin{pmatrix}
{\left[F\left(r\right)/r\right]\Phi_{\kappa m}}\\
{\left[iG\left(r\right)/r\right]\Phi_{- \kappa m}}
\end{pmatrix}
\tau ,
\end{equation}
where
\begin{equation}
\label{Eq2}
\Phi_{\kappa m} = \sum_{m_{l} m_{s}} {C^{l\ 1/2\ j}_{m_{l} m_{s} m}
Y_{l m_{l}}\left(\theta , \phi \right)\chi_{m_{s}} }   \; ,
\end{equation}
$j=\left|\kappa\right|-1/2$, and $l=\kappa$ for $\kappa>0$, but $l=-\left(\kappa+1\right)$ for $\kappa<0$ and $\tau$ indicates a proton or neutron. The regular and irregular Dirac Coulomb functions are generated as given by Young and Norrington \cite{YN94} employing the code COULCC \cite{T85}, and they are given the asymptotic form,
{\small\begin{align} 
F_{R} &= \sqrt{E+m}\; \sin \phi\left(r\right)  \  \ {\rm and} \  \ G_{R} =\sqrt{E-m} \; \cos \phi\left(r\right) \; ,  \nonumber\\
F_{IR} &=\sqrt{E+m}\; \cos \phi\left(r\right)  \  {\rm and}  \ \  G_{IR} =-\sqrt{E-m}\; \sin \phi\left(r\right) \; , \nonumber
\end{align}}
where $\phi \left(r\right) = kr + y \ln 2kr  + \delta^{\prime}_{\kappa} - l \pi /2 \;$, $k$ is the momentum of the proton in the center-of-momentum system, $y = Z e^{2} E/k$, $E^{2} = m^{2}_{p} + k^{2}$, $\delta^{'}_{\kappa} = \Psi - \arg\Gamma\left({\gamma + iy}\right) + {\frac{\pi}{2}} {\left(l+1-\gamma\right)}$, $e^{2i\Psi} = \frac{ie^{2}Z/k-\kappa}{\gamma+iy}$, and $\gamma = \left(\kappa^{2}-Z^{2} e^{4}\right)^1/2$. Throughout this paper, $c=\hbar = 1$.  Incoming and outgoing waves are constructed as
{\small\begin{equation}
F_{I} = F_{IR} - iF_{R}  \ {\rm and}  \  F_{O} = F_{IR} + iF_{R} \ {\rm making\; up}\; I_{c} , \nonumber\\
\end{equation}}
and
{\small\begin{equation}
G_{I} = G_{IR} - iG_{R}  \ {\rm and}   \ G_{O} = G_{IR} + iG_{R} \ {\rm making\; up}\; O_{c} ,\nonumber\\
\end{equation}}
where 
c indicates a particular channel, $\left|lj\kappa\tau ,J_{A}\left(J_{B}\right)  \right\rangle$, $J_{A}$ is the target spin, and $J_{B}$ the total angular momentum.  A wave function with unit outgoing flux is $O_{c}/\sqrt{2k_{c}}$.

The appropriate modifications for expanding the one-channel case, given in Ref. \cite{Ha09}, to the many-channel case are as follows.  The wave function is expanded within the channel radius as $\psi=\sum_{\lambda} A_{\lambda} \left|\lambda\right\rangle$. The set of $ \left|\lambda\right\rangle$ will be Dirac oscillators coupled to the spin of the target. The Hamiltonian to be solved is
{\small\begin{equation} 
\label{Eq3}
\sum_{\lambda^{\prime} = 1} {\left[\left\langle \lambda\left| H - E \right| \lambda^{\prime}\right\rangle + \sum_{c} {\gamma_{\lambda c} \left(b_{\lambda^{\prime} c} - b_{c}\right) \gamma_{\lambda^{\prime} c}} \right] A_{\lambda^{\prime}}} = 0 \: .
\end{equation}}
where
\begin{equation}
\label{Eq4}
b_{ c} = G_{c}\left(a_{c}\right) / F_{c}\left(a_{c}\right)  \; ,
\end{equation}
\begin{equation}
\label{Eq5}
b_{\lambda c} = G_{\lambda c}\left(a_{c}\right) / F_{\lambda c}\left(a_{c}\right)  \; ,
\end{equation}
and
\begin{equation}
\label{Eq6}
\gamma_{\lambda c} = F_{\lambda c}\left(a_{c}\right) \; .
\end{equation}
$G_{c}$ and $F_{c}$ are the components of the physical wave function in channel $c$.  The theory is placed in calculable form in the method of Philpott \cite{P75} in which one finds a transformation $T$ such that
{\small\begin{equation}
\label{Eq7}
\sum_{\lambda\lambda^{\prime}} {T_{\mu\lambda} \left[ \left\langle \lambda\left| H \right| \lambda^{\prime}\right\rangle +  \sum_{c} {\gamma_{\lambda c} b_{\lambda^{\prime} c} \gamma_{\lambda^{\prime} c}} \right] T_{\lambda^{\prime} \mu^{\prime}}} = E_{\mu} \delta_{\mu\mu^{\prime}} .
\end{equation}}
With this transformation, Eq. (\ref{Eq3}) becomes
\begin{equation}
\label{Eq8}
\sum_{\mu^{\prime}} {\left[ \left( E_{\mu} - E\right )\delta_{\mu\mu^{\prime}} -  \sum_{c} {\gamma_{\mu c} b_{c} \gamma_{\mu^{\prime} c}} \right] A_{\mu^{\prime}}} = 0 \; ,
\end{equation}
where $\gamma_{\mu c} = \sum_{\lambda} \gamma_{\lambda c} T_{\lambda\mu}$ and $A_{\lambda} = \sum_{\mu} T_{\lambda\mu}A_{\mu}$. One changes $c$ to $c^{\prime}$ in Eq. (\ref{Eq8}), multiplies by $\gamma_{\mu c}/\left(E_{\mu}-E\right)$, and sums over $\mu$ to obtain
\begin{equation}
\label{Eq9}
\gamma_{c} = \sum_{\mu c^{\prime}} \frac{ \gamma_{\mu c^{\prime}} \gamma_{\mu c} b_{ c^{\prime}}   }{E_{\mu}-E}   \sum_{\mu^{\prime}} {A_{\mu^{\prime}} \gamma_{\mu^{\prime} c^{\prime}}}  ,
\end{equation}
or
\begin{equation}
\label{Eq10}
\sum_{c^{\prime}} {\left[\delta_{cc^{\prime}} - R_{cc^{\prime}}b_{c^{\prime}}\right] \gamma_{c^{\prime}}} = 0  ,
\end{equation}
where
\begin{equation}
\label{Eq11}
\gamma_{c} = \sum_{\mu} {A_{\mu} \gamma_{\mu c}} ,
\end{equation}
and
\begin{equation}
\label{Eq12}
\ R_{cc^{\prime}} = \sum_{\mu} \gamma_{\mu c} \gamma_{\mu c^{\prime}}/ \left(E_{\mu} - E\right)  .
\end{equation}

The amplitudes are extracted from Eq. (\ref{Eq9}),
{\small\begin{equation}
\label{Eq13}
A_{\mu} = \frac{1}{E_{\mu}-E} \sum_{c} \gamma_{\mu c} b_{c} \gamma_{c}= \frac{1}{E_{\mu}-E} \sum_{c} \gamma_{\mu c} G_{c}\left(a_{c}\right)   .
\end{equation}}

A general solution to the coupled channels wave function in the external region is \cite{LaTh58}
\begin{equation}
\label{Eq14}
\Psi = \sum_{c} \left( \frac{x_{c}}{\sqrt{2k_{c}}} O_{c}+ \frac{y_{c}}{\sqrt{2k_{c}}} I_{c}  \right) \ .
\end{equation}

The collision matrix $\textbf{S}$ provides an expression for the $x_{c}$ in terms of the $y_{c}$.  In matrix notation
\begin{equation}
\label{Eq15}
\textbf{x}=-\textbf{S}\textbf{y}  .
\end{equation}

From Eqs. (\ref{Eq4}), (\ref{Eq6}), (\ref{Eq10}) and (\ref{Eq14}), the fundamental $R$ matrix equation for the relativistic case relates the upper components of the wave functions to the lower,
{\small\begin{align}
\label{Eq16}
F_{c^{\prime}}& = \sum_{c^{\prime}} R_{cc^{\prime}} G_{c^{\prime}}=  \sum_{c^{\prime}} R_{cc^{\prime}} \left[ G_{Oc^{\prime}}x_{c^{\prime}}/\sqrt{2k_{c^{\prime}}} + G_{Ic^{\prime}}y_{c^{\prime}}/\sqrt{2k_{c^{\prime}}}  \right] \nonumber\\
&= F_{Oc}x_{c}/\sqrt{2k_{c}} + F_{Ic}y_{c}/\sqrt{2k_{c}}  .
\end{align}}
If one defines diagonal matrices $v_{cc^{\prime}}=2k_{c}\delta_{cc^{\prime}}$, $x_{cc^{\prime}}=\delta_{cc^{\prime}} x_{c}$, $y_{cc^{\prime}}=\delta_{cc^{\prime}} y_{c}$, $G_{Occ^{\prime}}=\delta_{cc^{\prime}} G_{Oc}$, $G_{Icc^{\prime}}=\delta_{cc^{\prime}} G_{Ic}$, $F_{Occ^{\prime}}=\delta_{cc^{\prime}} F_{Oc}$ and $F_{Icc^{\prime}}=\delta_{cc^{\prime}} F_{Ic}$, this equation can be written as $ \textbf{F}_{O}\; \textbf{v}^{-\frac{1}{2}} \; \textbf{x}   + \textbf{F}_{I} \; \textbf{v}^{-\frac{1}{2}}\;\textbf{y} =\textbf{R}  \textbf{G}_{O}\; \textbf{v}^{-\frac{1}{2}} \;\textbf{x} + \textbf{R}  \textbf{G}_{I} \; \textbf{v}^{-\frac{1}{2}} \;\textbf{y} $. If one solves for $\textbf{x}$, one obtains the form in Eq. (\ref{Eq15}), $\textbf{x}=-\textbf{S}\textbf{y} $, where
\begin{equation}
\label{Eq17}
\textbf{S}=  \textbf{v}^{\frac{1}{2}}  \left[ \textbf{F}_{O}-\textbf{R}\textbf{G}_{O} \right]^{-1}      \left[ \textbf{F}_{I}-\textbf{R}\textbf{G}_{I} \right] \textbf{v}^{-\frac{1}{2}}  .
\end{equation}
Then the $\textbf{T}$ matrix, $T_{cc^{\prime}}$, is in the usual form, $i\left(\delta_{cc^{\prime}} -S_{cc^{\prime}}\right)/2$.

The scattering amplitude is found by following standard techniques.  Target (residual) states are noted as $\left|\alpha J_{A}M_{A} \right\rangle$, where $J_{A}$, $M_{A}$ are the spin and its projection and $\alpha$ distinguishes among states of the same spin.  Target states may be coupled to the angular momentum of the projectile yielding states with total angular momentum and projection $\left|\alpha J_{A} lj J_{B}M_{B} \right\rangle$. The scattering states are designated by the target state, its projection, and the spin projection of the projectile $\sigma$.  The resulting scattering amplitude is

{\small\begin{align}
\label{Eq18}
& \left\langle  f\right\rangle_{\alpha \sigma M_{A},\alpha^{\prime} \sigma^{\prime} M^{\prime}_{A}}  \nonumber\\
& =\frac{1}{k}  \sum \sqrt{4\pi\left(2l+1\right)} C^{l\: 1/2\ j}_{0 \: \sigma \ m} C^{J_{A}\ j\ J_{B}}_{M_{A}\ m\ M_{B}} C^{J^{\prime}_{A}\ j^{\prime}\ J_{B}}_{M^{\prime}_{A}\ m^{\prime}\ M_{B}} C^{l^{\prime}\ 1/2\ j^{\prime}}_{m^{\prime}_{l} \: \sigma \ m^{\prime}}   \nonumber\\
& \times i^{\left(l-l^{\prime}\right)} e^{i\left(\delta^{\prime}_{\kappa}+\delta^{\prime}_{\kappa^{\prime}}\right)} T_{\alpha J_{A} lj J_{B},\alpha^{\prime} J^{\prime}_{A} l^{\prime}j^{\prime} J^{\prime}_{B}}  .
\end{align}}
The sum is over $\kappa$, $\kappa^{\prime}$, $\alpha^{\prime}$, $J_{B}$, $M_{B}$, $m$, $m^{\prime}$, and $m^{\prime}_{l}$. Scattering observables can then be calculated from the scattering amplitude.  For instance, the cross section would be given by
\begin{align}
\label{Eq19}
& \frac{d\sigma}{d\Omega} \left(\theta\right)=\frac{1}{2\left(2J_{A}+1\right)} \sum_{\sigma\sigma^{\prime} M_{A} M^{\prime}_{A}} \left| \left\langle f_{c}\right\rangle_{\sigma\sigma^{\prime} \delta_{ J_{A} \alpha M_{A}, J^{\prime}_{A} \alpha^{\prime} M^{\prime}_{A} }}      \right.  \nonumber\\
& +\left. \left\langle  f\right\rangle_{\alpha \sigma M_{A},\alpha^{\prime} \sigma^{\prime} M^{\prime}_{A}} \right|  ,
\end{align}
where $\left\langle f_{c}\right\rangle_{\sigma\sigma^{\prime}}$ is the relativistic Coulomb scattering amplitude \cite{Ha09}, taken to be diagonal in the target states.

\section{Relativistic continuum shell model}

The random-phase approximation and TDA equations for QHD were derived in Ref. \cite{F85} following Ref. \cite{FW71} and appear the same as the nonrelativistic equations.  The TDA equation is 
{\small\begin{equation}
\label{Eq20}
\left({\epsilon_{\lambda} - \epsilon_{\mu} - \epsilon}\right) C_{\lambda\mu} + \sum_{\alpha\beta} \left[  \left\langle {\beta\lambda}\left|{V}\right|{\alpha\mu}\right\rangle - \left\langle {\beta\lambda}\left|{V}\right|{\mu\alpha}\right\rangle  \right] C_{\alpha\beta} = 0  .
\end{equation}}

To apply QHD to finite nuclei the meson fields are taken as classical fields and a set of Dirac equations solved in the Hartree approximation \cite{SW86,HMS91}. The $\sigma$ and $\omega$ coupling constants were fit to the saturation properties of equilibrium nuclear matter and the $\rho$ coupling constant determined from the bulk symmetry energy. The $\sigma$ mass was determined so as to reproduce the r.m.s. radius of $^{40}$Ca, and for the Coulomb potential, one uses the contribution to the baryon density of protons only, while for the $\rho$, one uses half the difference between the proton and neutron densities.  In order to implement the QHD results in a TDA equation, the classical meson fields are replaced with one-meson exchange potentials as in Ref. \cite{F85}. 
\begin{align}
\label{Eq21}
&  V = \frac{-g^{2}_{\sigma}}{4 \pi} \frac{e^{-m_{\sigma} r}}{r} + \gamma^{\lambda}_{1} \gamma_{2 \lambda} \frac{g^{2}_{\omega}}{4 \pi} \frac{e^{-m_{\omega} r}}{r}  \nonumber\\
& + \gamma^{\lambda}_{1} \gamma_{2 \lambda} \frac{\tau_{1}\cdot\tau_{2}}{4} \frac{g^{2}_{\rho}}{4 \pi} \frac{e^{-m_{\rho} r}}{r} + \gamma^{0}_{1} \gamma^{0}_{2} \frac{e^{2}}{r}   ,
\end{align}
where the Coulomb interaction has been included.  The coupling constants employed are the same as those from QHD calculations, although it is not clear that these should be appropriate in structure calculations.  The finite-Hartree (FH) coupling constants of Ref. \cite{SW86} are shown in Table \ref{Table1}.  In addition, the hole SPEs, $\epsilon_{\mu}$, and the wave functions are taken as those from the FH, QHD calculation, generated with the code TIMORA \cite{HMS91}.  (A nucleon mass is added to the actual output of the code to obtain $\epsilon_{\mu}$.)  However, the particle SPE, $\epsilon_{\lambda}$, are replaced by the interaction of the particle with the core nucleons,
{\small\begin{align}
\label{Eq22}
& E_{jj^{\prime}}= \left\langle j  \right| \bm{\alpha} \cdot \textbf{p} + m \bm{\beta}  \left| j^{\prime} \right\rangle  \nonumber\\ 
&+ \sum^{occ}_{j_{c},J} \frac{2J+1}{2j+1}  \left\langle \gamma_{0} j \gamma_{0} j_{c}\left(J\right) \: \right| V \left( \left|j^{\prime}j_{c}\left(J\right) \right\rangle  -\left(-\right)^{j^{\prime}+j_{c}-J} \left|j_{c}j^{\prime}\left(J\right)\right\rangle  \right) ,
\end{align}}
where the sum $j_{c}$ is over proton and neutron states below the Fermi surface.  The integrals extend only to the $R$ matrix radius. The notation is that $\left\langle  \gamma_{0} j\right|$ is $\bar{u}=u^{+} \gamma_{0}$ with angular momentum $j$.  A similar SPE definition could be made for the hole states with $\left|j\right\rangle$ and $\left|j^{\prime}\right\rangle$ replaced with $\left|j_{h}\right\rangle$ giving $E_{j_{h}j_{h}}$.
\begin{table}
\caption{\label{Table1} Coupling constants.  FH is finite Hartree; HF is Hartree-Fock.}
\begin{ruledtabular}
\begin{tabular}{cccc}
${Meson}$  & {$Mass$ (MeV)} & ${FH, g^{2}}$ & ${HF, g^{2}}$\\
\hline\\
$\sigma$  & ${520}$ & ${109.6}$ & ${89.6}$\\
$\omega$  & ${783}$ & ${190.4}$ & ${102.6}$\\
$\rho$  & ${770}$ & ${65.2}$ & ${12.4}$\\
$\pi$  & ${138}$ & ${0}$ & ${181}$\\
\end{tabular}
\end{ruledtabular}
\end{table}

 Eq. (\ref{Eq20}) is now an equation to be solved for the particle wave functions for a given energy.  The basis functions, the $\left|\lambda\right\rangle$ of Eq. (\ref{Eq3}), are particle-hole functions where the $\left|j\right\rangle$ are Dirac oscillators specified by $\left|nlj\kappa\right\rangle$ and hole states are the QHD states generated with parameters FH as used to construct the targets in Ref. \cite{Ha09}.  Hole states are the target states with spin $j_{h}=J_{A}$.  A matrix element of the Hamiltonian (excluding the Bloch operator) within the $R$ matrix radius is 
{\small\begin{align}
\label{Eq23}
& \left\langle j\otimes j_{h} \left(J_{B}\right)\left|H\right|j^{\prime}\otimes j_{h^{\prime}} \left(J_{B}\right)\right\rangle   \nonumber\\ 
& = E_{jj^{\prime}} - \epsilon_{j_{h}}\delta_{j_{h}j_{h^{\prime}}} - \sum_{J} \left(2J+1\right) W\left(jj_{h}j_{h^{\prime}}j^{\prime};J_{B}J\right) \nonumber\\ 
& \times \left\langle \gamma_{0} j_{h^{\prime}} \gamma_{0} j \left(J\right)\right| V \left[ \left|j_{h} j \left(J\right)\right\rangle - \left(-\right)^{j_{h}+j-J} \left|j j_{h} \left(J\right)\right\rangle \right] ,
\end{align}}
The particle wave functions are orthogonal to the hole states and the exchange terms are calculated exactly in the method of Ref \cite{Ha09}.

To check whether replacing the classical fields with one meson exchange potentials is appropriate, one can compare the single particle energies of the hole states calculated from QHD and those calculated by the interaction of the hole state with particles in the core $E_{j_{h}j_{h}}$. The comparison is made for two nuclei, $^{16}$O and $^{90}$Zr. One is interested in $^{16}$O, because it is the subject of numerous $\left(e,e^{\prime}x\right)$ experiments and the question of the role of relativity in these reactions, however, only six SPEs can be compared for this nucleus. Therefore, a comparison is first made for $^{90}$Zr which has 21 SPEs.  The $^{90}$Zr comparison is shown in Table \ref{Table2} for $E^{x}_{jj^{\prime}}=E_{jj^{\prime}}-M_{N}$. The first column lists the QHD output from TIMORA.  The second column is from the one-pion exchange calculation with the same coupling constants.  Although the SPEs calculated with the potential are shifted upward slightly and have some difficulties with the spin-orbit splitting, the agreement between the two calculations is surprising.  In Table \ref{Table3} is shown the SPE comparison for $^{16}$O where the agreement is similar.  Also shown in this table are the experimental SPEs and those from a recent non-relativistic Hartree-Fock \cite{D11} calculation.  These last two columns demonstrate that the original QHD,FH calculation has some difficulty with the spin-orbit splitting.
\begin{table*}
\caption{\label{Table2} Single particle energies for $^{90}$Zr in MeV.  Column 1 is the QHD finite Hartree result; column 2 is the one-meson exchange with finite Hartree parameters; column3 is the same as column 2 plus pseudoscalar pions; column 4 is the same as column 2 plus pseudovector pions in the effective mass approximation; column 5 is the one-meson exchange with QHD Hartree-Fock parameters plus pseudoscalar pions.}
\begin{ruledtabular}
\begin{tabular}{cccccc}
${State}$ & {FH, QHD} & {FH, $E^{x}_{jj^{\prime}}$} & {FH, $E^{x}_{jj^{\prime}}$ ps} & {FH, $E^{x}_{jj^{\prime}}$ $M^{\*}$} & {HF, $E^{x}_{jj^{\prime}}$ ps} \\
\hline\\
${0s1/2\left(p\right)}$ & ${-52.43}$ & ${-49.11}$ & ${37.17}$ & ${-30.32}$ & ${-16.59}$ \\
${0p3/2\left(p\right)}$ & ${-42.06}$ & ${-40.32}$ & ${36.42}$ & ${-20.99}$ & ${-14.24}$ \\
${0p1/2\left(p\right)}$ & ${-39.54}$ & ${-35.30}$ & ${47.47}$ & ${-19.04}$ & ${-6.96}$ \\
${0d5/2\left(p\right)}$ & ${-30.15}$ & ${-29.96}$ & ${36.26}$ & ${-10.33}$ & ${-9.99}$ \\
${1s1/2\left(p\right)}$ & ${-20.93}$ & ${-16.25}$ & ${52.86}$ & ${-2.60}$ & ${3.22}$ \\
${0d3/2\left(p\right)}$ & ${-24.86}$ & ${-19.88}$ & ${54.27}$ & ${-6.56}$ & ${1.39}$ \\
${0f7/2\left(p\right)}$ & ${-17.62}$ & ${-18.73}$ & ${36.54}$ & ${0.52}$ & ${-4.04}$ \\
${1p3/2\left(p\right)}$ & ${-6.92}$ & ${-2.54}$ & ${52.25}$ & ${7.92}$ & ${10.96}$ \\
${0f5/2\left(p\right)}$ & ${-9.55}$ & ${-4.13}$ & ${57.14}$ & ${5.99}$ & ${9.17}$ \\
${1p1/2\left(p\right)}$ & ${-5.11}$ & ${0.10}$ & ${53.94}$ & ${8.96}$ & ${13.02}$ \\
${0s1/2\left(n\right)}$ & ${-62.72}$ & ${-54.55}$ & ${30.44}$ & ${-38.20}$ & ${-26.78}$ \\
${0p3/2\left(n\right)}$ & ${-51.04}$ & ${-44.17}$ & ${30.79}$ & ${-27.80}$ & ${-23.70}$ \\
${0p1/2\left(n\right)}$ & ${-48.80}$ & ${-39.92}$ & ${42.77}$ & ${-25.41}$ & ${-15.31}$ \\
${0d5/2\left(n\right)}$ & ${-38.18}$ & ${-32.82}$ & ${30.52}$ & ${-16.53}$ & ${-19.40}$ \\
${1s1/2\left(n\right)}$ & ${-30.30}$ & ${-21.19}$ & ${45.05}$ & ${-9.59}$ & ${-7.33}$ \\
${0d3/2\left(n\right)}$ & ${-33.20}$ & ${-23.53}$ & ${51.24}$ & ${-11.64}$ & ${-5.49}$ \\
${0f7/2\left(n\right)}$ & ${-25.07}$ & ${-21.34}$ & ${30.02}$ & ${-5.60}$ & ${-13.63}$ \\
${1p3/2\left(n\right)}$ & ${-15.94}$ & ${-7.84}$ & ${44.53}$ & ${0.73}$ & ${0.65}$ \\
${0f5/2\left(n\right)}$ & ${-17.23}$ & ${-7.22}$ & ${54.56}$ & ${1.67}$ & ${3.03}$ \\
${1p1/2\left(n\right)}$ & ${-14.03}$ & ${-4.94}$ & ${47.39}$ & ${2.33}$ & ${3.58}$ \\
${0g9/2\left(n\right)}$ & ${-12.26}$ & ${-10.08}$ & ${29.53}$ & ${4.27}$ & ${-6.47}$ \\
\end{tabular}
\end{ruledtabular}
\end{table*}
\begin{table}
\caption{\label{Table3} Single particle energies for $^{16}$O in MeV.  Column 1 is the QHD finite Hartree result; column 2 is the one-meson exchange with finite Hartree parameters; column 3 shows experimental values; column 4 shows results of a non-relativistic Hartree-Fock calculation.}
\begin{ruledtabular}
\begin{tabular}{ccccc}
${State}$ & {FH, QHD} & {FH, $E^{x}_{jj^{\prime}}$} & {Exp.} & {Ref. \cite{D11}} \\
\hline\\
${1s1/2\left(p\right)}$ & ${-37.2}$ & ${-36.2}$ & ${-37 \pm 4}$ & ${-35.4}$  \\
${1p3/2\left(p\right)}$ & ${-16.7}$ & ${-17.8}$ & ${-17.4}$ & ${-18.6}$  \\
${1p1/2\left(p\right)}$ & ${-8.8}$ & ${-4.0}$ & ${-12.1}$ & ${-12.5}$ \\
${1s1/2\left(n\right)}$ & ${-41.4}$ & ${-39.0}$ & ${-40 \pm 4}$ & ${-38.6}$ \\
${1p3/2\left(n\right)}$ & ${-20.6}$ & ${-20.8}$ & ${-21.8}$ & ${-21.8}$  \\
${1p1/2\left(n\right)}$ & ${-12.5}$ & ${-6.7}$ & ${-15.7}$ & ${-15.6}$  \\
\end{tabular}
\end{ruledtabular}
\end{table}

\begin{figure}[ht]
\includegraphics[width=8cm]{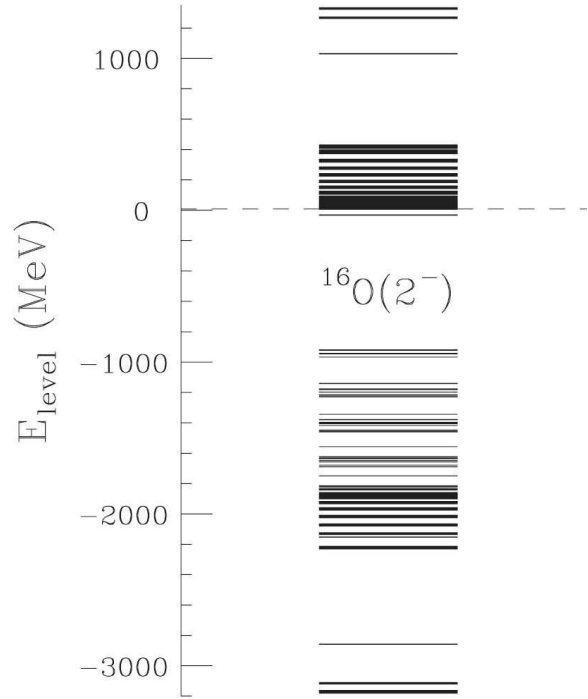}
\caption{\label{Fig1} $R$ matrix levels for $J^{\pi} = 2^{–}$.}
\end{figure}

The $R$ matrix is now calculated for $^{16}$O , and the $R$ matrix level energies for $J^{\pi}=2^{-}$ are plotted in Fig. \ref{Fig1}.  In a non-relativistic calculation, one would have $R$ matrix levels below threshold corresponding to bound states, levels above threshold corresponding to resonances, and levels very much above threshold that comprise the continuum.  These levels appear in the relativistic calculation as well, however, a nearly equal number of negative energy levels appear approximately one nucleon mass below threshold.  These levels are absolutely necessary for the cross section calculations.

\begin{figure}[ht]
\includegraphics[width=8cm]{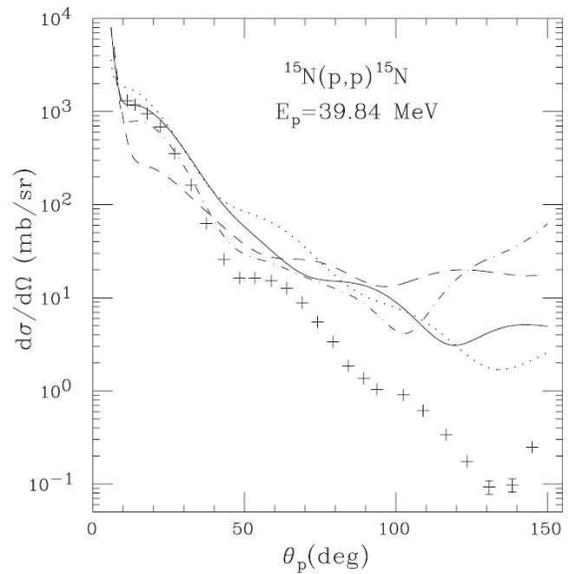}
\caption{\label{Fig2} Cross section for 39.84 MeV protons on $^{15}$N.  Solid line uses FH parameters; dashed line uses FH parameters plus pseudoscalar $\pi N$ coupling; dot-dashed line uses effective mass approximation to pseudovector $\pi N$ coupling; dotted line is non-relativistic calculation.  Data are from Ref. \cite{FW71}.}
\end{figure}

In Fig. \ref{Fig2} the solid line represents the calculated 39.84 MeV elastic scattering cross section for protons on $^{15}$N with the FH parameter set, the same set used to calculate the bound state wave functions and bound state SPEs.  Hole states $j_{h}$ are limited to the p-shell and their energies are taken as those from the QHD calculation.  No pions are included in the FH interaction.  The agreement with the data \cite{S69} is again surprisingly good.  With only four core states, no absorption, and such a simple interaction, one does not expect the calculation to fit the back-angle data.  However, based on equivalent non-relativistic calculations one does expect a reasonable fit to forward angles, and this is accomplished.  Included in Fig. \ref{Fig2} as a dotted line is the equivalent non-relativistic calculation with the recoil corrected continuum shell model \cite{HP80} and the M3Y \cite{B77} interaction.  The agreement with data is about the same.  The relativistic continuum shell model with the FH interaction is, therefore, a simple, relatively self-consistent, physically justified model in which to investigate relativistic contributions in knockout reactions.

\section{pions}

If one looks at the solid line in Fig. \ref{Fig2} as if it were scattering from a Woods-Saxon potential, one would consider altering the diffuseness to obtain a better fit.  Indeed, QHD, FH has no pions whose longer range would alter the surface properties of an equivalent Woods-Saxon.  Therefore a brief look at the possible role of pions is worthwhile.  The pions can be added by making the instantaneous approximation to the one-meson exchange propagator.  In the case of pseudoscalar coupling the energy transferred is set to zero and $\Delta\left(x^{\prime}-x\right) = \int \frac{d^{4}q}{\left(2\pi\right)^{4}} \frac{e^{-iq\left(x^{\prime}-x\right)}}{k^{2}-m^{2}_{\pi}+i\epsilon}$ becomes $-\int \frac{d^{3}q}{\left(2\pi\right)^{3}} e^{i\vec{k}\left(\vec{x}-\vec{x}^{\prime}\right)} /\left( \vec{k}^{2}+m^{2}_{\pi}-i\epsilon\right) \delta\left(t-t^{\prime}\right)$. The Fourier transform yields a Yukawa and the term $\gamma^{5}_{1} \gamma^{5}_{2} \tau_{1}\cdot\tau_{2} \frac{g^{2}_{\pi}}{4 \pi} \frac{e^{-m_{\pi} r}}{r}$ is added to Eq. (\ref{Eq21}).  This approximation looks quite adequate if the energy transferred is reasonably small.  However, the procedure is less satisfactory in the case of pseudovector coupling.  The vertex function $\left(f_{\pi}/m_{\pi}\right)\gamma_{5}\slashed{q}\tau_{i}$ includes a term with the energy transferred.  Setting this term to zero is very different than setting it to zero in the denominator of the propagator.  However, a simple approximation was proposed in Ref. \cite{F85} in which pseudovector coupling is approximated by using an effective nucleon mass in the pseudoscalar matrix elements.  The prescription is that one uses the Yukawa interaction above, multiplied by $[M^{*}(\textbf{x}_{1})/M][M^{*}(\textbf{x}_{2})/M]$, where $M^{*}=M-g_{s} \varphi (r)$ and $g_{s} \varphi (r)$  is the scalar potential for the hole states.  The effective mass approximation provides a density-dependence to the interaction, although a severe one, the interior pion potential being reduced by approximately 75$\%$. 

The $^{15}$N$(p,p)^{15}$N cross section with the FH coupling constants and pseudoscalar coupling with $g^{2}_{\pi}=181$ appears as a dashed line in Fig. \ref{Fig2}.  The cross section is more diffractive, but the fit is poor. This is reflected in the enormous increase in the $^{90}$Zr single particle energies as shown in column 3 of Table \ref{Table2}.  This increase shows that, for these coupling constants, the pions are producing repulsion, just as they did in the Hartree-Fock calculations of QHD \cite{SW86}.  Also shown in Fig. \ref{Fig2} as a dot-dashed line is the cross section with the effective mass approximation to pseudovector coupling.  This addition has a smaller effect, as one would expect from derivative coupling.  It also improves the diffraction peak locations, but the severity of the density dependence produces unusual back-angle behavior.  The effective mass approximation produces the SPEs in column 4 of Table \ref{Table2}, and they are certainly an improvement over the full pseudoscalar results.  It is now clear why the structure calculations in Ref. \cite{F85} preferred this approximation to pseudovector coupling over pseudoscalar coupling.  

\begin{figure}[ht]
\includegraphics[width=8cm]{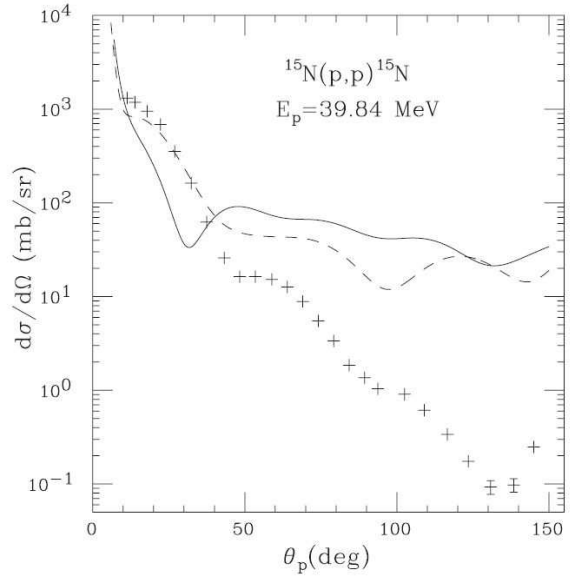}
\caption{\label{Fig3} Cross section for 39.84 MeV protons on $^{15}$N.  Dashed line uses HF parameters plus pseudoscalar $\pi N$ coupling; solid line is the same, but without pions.  Data are from Ref. \cite{FW71}.}
\end{figure}

Additional sets of coupling constants were obtained in the Hartree-Fock calculations of QHD which included pions in the coupling constant fit \cite{SW86}.  The $\pi N$ coupling was pseudovector.  One set of these coupling constants is given in Table \ref{Table1} under the title HF.  The SPE results of this representative set are shown in column 5 in Table \ref{Table2} for pseudoscalar coupling.  These SPEs are underbound.  Use of the effective mass approximation to pseudovector coupling produced extremely underbound SPEs and is considered unacceptable.  The cross section with pseudoscalar coupling is shown as the dashed line in Fig. \ref{Fig3} and with no pions as a solid line.  The inclusion of the pions improves the cross section considerably, especially the location of the diffraction peaks, but the cross section still rises at back angles.

One can conclude that pions have the capability to alter the cross section, especially in a manner normally associated with surface effects.  However, it would be very beneficial to have a better approximation for calculating the pseudovector matrix elements to adequately judge its effect.

\section{Conclusion}

This article provided the final derivations for an $R$ matrix formalism so that the many-coupled channels problem may be solved for systems in which binary breakup channels satisfy a relative Dirac equation.  Expressions for the collision matrix and the scattering amplitude are presented, and from these, one may calculate scattering observables.  In addition to providing scattering states, this $R$ matrix formalism may also be used to calculate resonances and bound states. 

The formalism is applied to relativistic continuum TDA calculations for $^{16}$O with interactions derived from relativistic mean field theory.  It was determined that even the simple $\sigma + \omega + \rho$ exchange with QHD coupling constants provides reasonable agreement with experimentally determined SPEs and the experimental $^{15}$N$(p,p)^{15}$N cross section at 39.84 MeV.  This is, therefore, a simple, physically justifiable interaction for later use in knockout reactions.  

\section*{Acknowledgments}
This work was supported by the National Science Foundation under grant PHY-0855339.

\end{document}